\documentstyle{lamuphys}
\input psfig
\input epsf

\makeatletter
\let\chapter\hid@chapter
\makeatother
\begin{document}
\pagenumbering{arabic}
\title{The hosts of $z=2$ QSOs}

\author{Itziar\, Aretxaga\inst{1}, B.J.\, Boyle\inst{2} and
Roberto J.\, Terlevich\inst{3}}

\institute{
Max-Planck Institut f\"ur Astrophysik, Karl Schwarzschildstr 1,
Postfach 1523, 85740 Garching, Germany
\and
Anglo-Australian Observatory, PO Box 296, Epping, NSW 2121 Australia
\and 
Royal Greenwich Observatory, Madingley Road, Cambridge CB3 0EZ, UK 
}

\maketitle

\def\lsim{\mathrel{\lower2.5pt\vbox{\lineskip=0pt\baselineskip=0pt
           \hbox{$<$}\hbox{$\sim$}}}}
\def\gsim{\mathrel{\lower2.5pt\vbox{\lineskip=0pt\baselineskip=0pt
           \hbox{$>$}\hbox{$\sim$}}}}
\newcommand{\itzifig}[6]{
    \epsfxsize=#1\epsffile[#2 #3 #4 #5]{#6}
     }
\newcommand{\cidfig}[6]{
    \protect\centerline{
    \epsfxsize=#1\epsffile[#2 #3 #4 #5]{#6}
     }}
\def\unilum{\hbox{erg s$^{-1}$}}
\def\q0{\hbox{q$_{\mbox{\scriptsize 0}}$}}
\def\H0{\hbox{H$_{\mbox{\scriptsize 0}}$}}
\def\Ho50{\hbox{H$_{\mbox{\scriptsize 0}} = 50$~\uniHo \ }}
\def\qo0p5{\hbox{q$_{\mbox{\scriptsize 0}} = 0.5$}}
\def\L0{\hbox{$\Lambda = 0$}}
\def\ojo{\fbox{\bf !`$\odot$j$\odot$!}}    
\def\uniHo{\hbox{Km s$^{-1}$ Mpc$^{-1}$}}

\begin{abstract}
We present results of the hosts of four high-redshift ($z \approx 2$) and
high luminosity ($M_B \lsim -28$~mag) QSOs, three radio-quiet one radio-loud, 
imaged in $R$ and $K$ bands.
The extensions to the nuclear unresolved source 
are most likely due to the hosts galaxies of these QSOs, with 
luminosities at rest-frame 2300\AA\ of at least 
$3-7\,$per cent of the QSO luminosity, and most
likely around $6-18\,$per cent of the QSO luminosity.  
Our observations show that, if the extensions we have 
detected are indeed galaxies, extraordinary big and luminous 
host galaxies are not only a characteristic of radio-loud objects, but of 
QSOs as an entire class.
\end{abstract}

\section{Introduction}

The study of high redshift ($z \approx 2$) QSOs offers a unique opportunity 
to investigate conditions in the early universe. In the currently favoured 
cold dark matter cosmogony, this epoch corresponds to the period when normal 
galaxies formed through hierarchical coalescence (Carlberg 1990), thereby 
giving rise to enormous concentrations of gas in the center of the galaxies,
which could feed a central black hole (Haehnelt \& Rees 1993) or provoke 
a massive 
starburst episode (Terlevich \& Boyle 1993). As such, this picture is 
consistent with the observation that the QSO population peaks 
at these redshifts (Schmidt et al. 1991). 

Searches for galaxies hosting high-redshift QSOs were first carried
out in radio-loud objects, with spectacular results:
luminosities several magnitudes brighter than the most luminous
galaxies in the nearby Universe were observed (Lehnert et al. 1992).
However, 
radio-loud quasars are only a small fraction ($<1$\%) of all QSOs, and 
many of the 
conclusions drawn from radio-loud objects might be 
unrepresentative of the conditions in the early universe. 
It is, therefore, important to look into the properties of radio-quiet QSOs,
as they may be better indicators of the general properties of 
galaxies at high redshifts. 
We present here deep multicolor images of one high-redshift radio-loud
and three high-redshift radio-quiet QSOs.

\section{Data analysis}

The four QSOs were observed 
in the Harris $R$ passband 
with the auxiliary port of the 4.2m William Herschel Telescope (WHT) 
at the Observatorio de Roque de los Muchachos in La Palma, and in $K$
band with the Cassegrain focus of the 3.5m at the Observatorio de Calar 
Alto in Almer\'{\i}a. 
The QSOs were
selected so as to have bright stars in the field
($20 \lsim \theta \lsim 50$~arcsec), which enabled us to 
define the point spread function (PSF)
of each observation accurately (see Aretxaga et al. 1995,1997 for 
a detailed description of data and analysis). 

For each QSO field, we defined
a 2-dimensional PSF using the brightest of the closest stellar 
companion to the QSO.
We then scaled the PSF to match the 
luminosity of the QSO and other nearby stars over the same region, and 
subtracted the scaled PSF from them.  The remaining residuals in the 
non-PSF stars provided an accurate check of the validity of the subtraction
process.
We accepted the PSF subtraction if the residuals in the non-PSF stars
accounted for less than 1$\sigma$ of the Poisson noise expected from the 
subtraction technique.
We have detected $R$-band residuals in excess of 
3$\sigma$ in the following QSOs:
1630.5$+$3749 (4$\sigma$), PKS~2134$+$008 (3$\sigma$) and Q~2244$-$0105 
(3.7$\sigma$). 
All the residuals show a `doughnut' shape with a well of 
negative counts in the centre. This indicates that there is a 
flatter component below the PSF in the centres of the QSOs, from which the 
nuclear (PSF) contribution has been over-subtracted.
As an example, Fig.1 shows
the $R$-band residuals for the 
radio-quiet QSOs 1630.5$+$3749, after subtracting a luminosity-scaled
PSF. 

\begin{figure*}
\hspace{0.5truein}
\hbox{\psfig{{figure=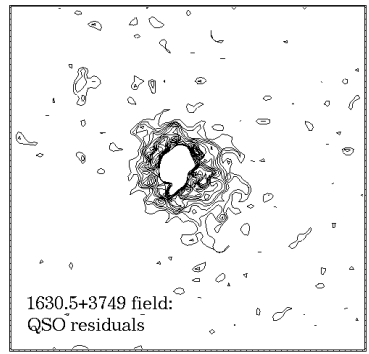,width=3.truein,height=2.3truein,rwidth=2.truein,rheight=2.truein}}}
\caption{$R$-band residuals of the QSO 1630.5$+$3749 after PSF subtraction
(Aretxaga et al. 1995). The size of the frame is $6''$x$6''$ .}
\end{figure*}

To estimate 
the true luminosity of these systems, we subtracted smaller amounts of 
the PSF in order either 
a) to produce zero counts in the center of the residuals
or b) to achieve a flat-top 
profile with no depression in the center.
We regard 
these quantities as lower limits (3--7\% of the QSO luminosity)
and best estimates (6--18\% of the QSO luminosity), respectively, of 
the total luminosity
of these extended components (Table~1). 
In all cases, the FWHM of the flat-top residual profiles are 
significantly larger than the FWHM of the stars in each field:
$1''.05$ vs. $0''.7$ for 1630.5$+$3749, $0''.8$ vs. $0''.7$ for PKS 2134$+$008 
and $0''.84$ vs. $0''.7$ for Q~2244$-$0105.

The $K$-band images of the same hosts show no significant extensions 
to the stellar profiles (see Fig.2), over a $1\sigma$ limiting magnitude 
$\mu_K \approx 23$~mag/arcsec$^2$ (Aretxaga et al. 1997). These limits
are consistent with previous non-detections of $z \approx 2$ radio-quiet 
hosts in $K$ band
(Lowenthal et al. 1995). If there are no
colour gradients, from the $r\sim 2-3''$ non-detection limits, 
the colours of the hosts are $R-K \lsim 3.3$~mag.

\vspace*{-1cm}
\begin{table*}
\begin{center}
\caption{Magnitudes of the QSOs and their extensions (Aretxaga et al. 1995)}
\begin{tabular}{lcccccc}
 Name & $M_R^*$ (QSO) & $M_R^{1}$  & $R^1$ & $M_R^2$ & $R^2$ & FWHM$^2$\\
      &  &  & & &  &(arcsec)\\
1630.5$+$3749   & $-28.7$ & $-24.8$ & $21.7$ &  
$-25.9$ & $20.9$   &  1.05\\
PKS 2134$+$008  & $-30.1$ & $ -25.9$ & $20.4$ &  
$-26.6$    & $19.8$ &  0.80  \\
Q 2244$-$0105   & $-29.0$ & $ -25.6$ & $20.9$ &  
$-26.7$ & $19.9$   &  0.84  \\
\end{tabular}  
\end{center}    
{\footnotesize
$^*$ Total QSO $M_R$ absolute magnitude, including host (\Ho50, \qo0p5).\\
$^1$ Properties for hosts with zero counts in the center after the PSF 
subtraction.\\
$^2$ Properties for hosts with flat-topped profile after the PSF subtraction
.}
\end{table*}    

\vspace*{-0.5cm}
\begin{figure*}
    \cidfig{2.8in}{305}{179}{574}{432}{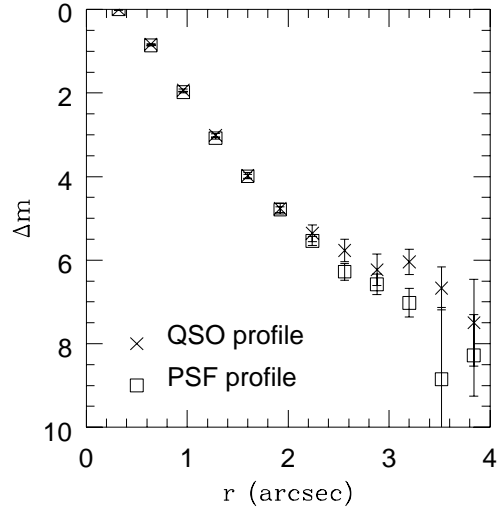}
\caption{Comparison between the QSO (1630.5$+$3749)
radial profile and  the PSF stellar radial profile in
$K$-band, where crosses indicate QSO and squares stellar profiles}
\end{figure*}

\section{Discussion: the origin of the 'fuzz'}

From a sample of one radio-loud and three radio-quiet QSOs with
suitable PSF stars, we have detected $R$-band extensions in three
cases (Aretxaga et al. 1995) and no extensions in $K$-band (Aretxaga et al. 
1997). The best estimates for $R$-band (2300\AA\ rest-frame) 
luminosities of these systems  lie between
6--18\% of the total QSO luminosity.  

\subsection{Scattering}

For the lobe-dominated radio-loud QSO sample of Lehnert et al. (1992), 
the red colours of the hosts 
favour a host galaxy origin of the excess light, rather
than scattering by dust or electrons in the halo of the QSOs. 
Our colour limits for the excess light, $R-K \lsim 3.3$~mag, include
colours as blue as those of the QSOs themselves ($R-K \approx 2.3$~mag), 
and are therefore
consistent with the colours expected from the optically 
thin scattering case and, also, 
with those of a young stellar population.  
They, however, exclude the colours of passively evolved
bulge populations (eg. Bressan et al. 1993).

However, most scattering models proposed to date 
require the presence of a powerful transverse 
radio-jet (e.g. Fabian 1989) which is unlikely
to be present in either the radio-quiet QSOs, or the
core-dominated radio-loud QSO 
around which we have detected extensions.

\subsection{Nebular continuum}

An alternative origin for the hosts could be extended nebular continuum, 
seen to be a major contribution to the UV continuum in three powerful 
radio galaxies (Dikson et al. 1995).
However, if our hosts are due to nebular luminosities
of $M_R\sim-26.5$~mag,
the predicted narrow H$\beta$
luminosities would be about $3 \times 10^{44}$~\unilum.
From the PSF light we derive that the luminosity of the broad
component is about double that. If this is so, 
the QSOs would exhibit prominent narrow lines with
central peak intensities of more than 3 times those of the broad lines. 
The spectra of our QSOs do not show such prominent narrow lines .

\subsection{Host galaxies}

  There is some circumstantial evidence that the extensions we have detected
are most probably the galaxies which host these QSOs:

a) The radial profile of 
the $R$-band hosts, derived from the flat-top solutions, 
falls approximately as an $r^{1/4}$-law for radii $r \gsim 0.6$~arcsec
(Fig. 3).
Profiles derived for radii smaller than the FWHM of the observations 
are usually unreliably 
recovered by flat-top subtractions, as confirmed by our the numerical
simulations of galaxy$+$PSF. However, total
luminosities 
and sizes are parameters which can be recovered well 
if the galaxy contribution 
exceeds 3\%\ of the QSO$+$galaxy system.

\begin{figure*}
    \cidfig{2.8in}{25}{420}{298}{699}{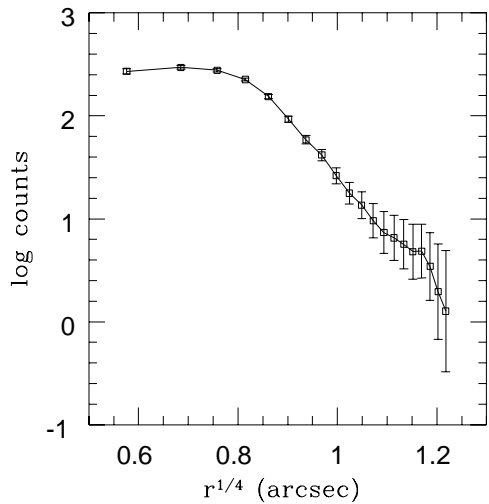}
\caption{ Radial profile of the $R$-band host of the QSO 1630.5$+$3749,
in a log counts vs. $r^{1/4}$ diagram: an $r^{1/4}$ profile
would appear as a straight line.
}
\end{figure*}

\begin{figure*}
    \cidfig{2.8in}{24}{154}{293}{426}{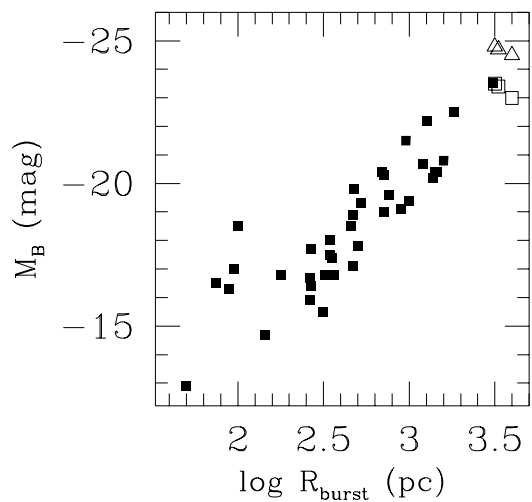}
\caption{Luminosity--size relationship for nearby H~II galaxies
(Telles 1995). H~II galaxies are marked with filled squares.
The three QSOs studied here lie in this relation 
if their SEDs are $f_\nu \propto \nu^0$ (open triangles) to 
$f_\nu \propto \nu^{0.5}$ (open squares). These SEDs are 
typical of young H~II galaxies}
\end{figure*}

b) The luminosities and radii of our hosts lie in the luminosity-radius
relation of local young H~II galaxies (Fig.4). We converted the UV luminosities
of the hosts (observed $R$-band) to rest-frame $B$-band
using the  spectral energy 
distribution (SED) of local H~II galaxies 
($f_\nu \propto \nu^\alpha$, with $0 \lsim \alpha \lsim 0.5$).
This is equivalent to converting the $B$-band luminosities of 
the H~II galaxies to rest-frame 2300~\AA\, and then comparing the 
UV luminosity-radius relationship of H~II galaxies with that of the hosts.
Notice that there is
at least one H~II galaxy which is as big and luminous as our hosts.
At $z\approx 2$, an unevolved $L_\star$ galaxy with SED typical of 
an H~II galaxy would appear to be about 3 mag fainter than the hosts we have
detected. The star formation rates involved would be 
of the order of a few hundreds of solar masses per year.

Galaxies as luminous as the extensions detected here
have already been found in the imaging survey of lobe-dominated 
radio-loud QSOs carried out by Lehnert et al. (1992). 
Four of the objects of their sample,
with similar redshifts to those in our sample, 
show `fuzz' around the PSF of the nucleus. In the observed $R$ frame
the absolute magnitude of this `fuzz' ranges from
$-25.6$ to $-26.9$~mag,
as derived from the $B$ and $K$ colours they report, which compares 
to the $-25.9$ to $-26.7$~mag we found in our study of two radio-quiet and one
core-dominated radio-loud QSO.
Our observations show that, if the hosts we have 
detected are indeed galaxies, young massive and luminous 
galaxies are not only a characteristic of radio-loud QSOs, but of 
QSOs as an entire class. 
Indeed, the only radio-loud QSO studied in this sample does not exhibit a 
significantly larger or a more luminous extension than those of radio-quiet
QSOs.  One of the radio-quiet QSOs exhibits no significant
evidence for any extension in any band.

\vspace*{0.3cm}
\noindent
{\bf Acknowledgments}: 
This work was supported in part by the 'Formation and Evolution of Galaxies'
network set up by the European Commission under contract ERB FMRX-CT96-086
of its TMR programme. IA also has been partly supported
by the EEC HCM fellowship ERBCHBICT941023. 
%

%
%


\begin{thebibliography}
%
%
  \bibitem{}{}{} Aretxaga I., Boyle B.J., Terlevich R.J. 1995, 
MNRAS, 275, L27.
  \bibitem{}{}{} Aretxaga I., Boyle B.J., Terlevich R.J. 1997, 
MNRAS, in preparation.
 \bibitem{}{}{} Carlberg R., 1990, ApJ, 350, 505.
 \bibitem{}{}{} Bressan et al. 1994, ApJS,94,63. 
 \bibitem{}{}{} Dickson R., Tadhunter C., Shaw M., Clark, N \& Morganti R.,
1995, MNRAS, 273, L29.
 \bibitem{}{}{} Fabian, A.C. 1989, MNRAS, 238, 41P.
 \bibitem{}{}{} Haehnelt M.G. \& Rees, M., 1993, MNRAS, 263,168.
 \bibitem{}{}{} Lehnert M.D, Heckman T., Chambers K.C. \& 
Miley G.K., 1992, ApJ, 393, 68.
 \bibitem{}{}{} Lowenthal J.D., Heckman T.M., Lehnert M.D. 
\& Elias J.H., 1995, ApJ, 439, 588.
 \bibitem{}{}{} Schmidt M., Schneider D.P. \& Gunn J.E., 1991, in 
ASP Conf. Ser. 21, p109.
  \bibitem{}{}{} Telles E. 1995, Ph.D. Thesis, Cambridge Univ.
 \bibitem{}{}{} Terlevich R.J. \& Boyle, B.J., 1993, MNRAS, 262, 491.
\end{thebibliography}
\end{document}